# The QSO variability-luminosity-redshift relation

Roberto Cid Fernandes Jr.[1,2,3], Itziar Aretxaga[2] and Roberto Terlevich[2]
[1] *Departamento de Física, CFM, UFSC. Campus Universitário, Trindade, Caixa Postal 476, 88040-900 Florianópolis, SC, Brazil*
[2] *Royal Greenwich Observatory, Madingley Road, CB3 0EZ, Cambridge*
[3] *Institute of Astronomy, Madingley Road, CB3 0HA, Cambridge*

## ABSTRACT

The relationship between variability, luminosity and redshift in the South Galactic Pole QSO sample is examined in an effort to disentangle the effects of luminosity and redshift in the amplitude of the optical variations. The anticorrelation between variability and luminosity found by other authors is confirmed. Our analysis also supports claims that variability increases with redshift, most likely due to an anticorrelation between variability and wavelength. In particular, our parametric fits show that the QSO variability-wavelength relation is consistent with that observed in low-luminosity nearby active galactic nuclei.

The results are used to constrain Poissonian-type models. We find that if QSO variability results from a random superposition of pulses, then the individual events must have B-band energies between $\sim 10^{50}$ and a few times $10^{51}$ erg and time-scales of $\sim 2$ yr. Generalized Poissonian models in which the pulse energy and lifetime scale with luminosity are also discussed.

## 1 INTRODUCTION

Variability is a conspicuous property of Active Galactic Nuclei (AGN) over a wide range of luminosities (from LINERs to QSOs) and redshifts (from essentially 0 up to $z > 4$). Given that the emissivity of QSOs peaks in the optical-UV range, creating the so-called Big UV Bump, the study of the variability properties in this spectral range is central to the understanding of the energy-generation mechanism.

Determining the way in which the amplitude of the variations depends on the source luminosity and redshift would be an important step towards constraining AGN models. Several observational studies have addressed this issue by using long term monitoring campaigns on large samples of objects, mostly carried out with photographic plates (Angione 1973, Uomoto, Wills & Wills 1976, Bonoli et al. 1979, Trèvese et al. 1989, Netzer & Sheffer 1983, Pica & Smith 1983, Cristiani, Vio & Andreani 1990, Giallongo, Trèvese & Vagnetti 1991, Hook et al. 1994, Trèvese et al. 1994, Cristiani et al. 1996, Di Clemente et al. 1996). Despite some early disagreement, it is now firmly established that there is an anticorrelation between variability and luminosity, such that the relative amplitude of the variations decreases as we move up the luminosity scale (e.g. Hook et al. 1994, Cristiani et al. 1996).

Given the strong correlation between luminosity and redshift in most samples, the relationship between variability and redshift is not easily disentangled from that between variability and luminosity. Cristiani et al. (1996) found a significant positive correlation between variability and redshift based on the analysis of 486 objects merged from three different samples, in agreement with a similar trend previously reported by Giallongo et al. (1991). The latter authors, however, did not find any significant correlation between variability and luminosity, whereas Cristiani et al. did. A strong anticorrelation between variability and luminosity was also found by Hook et al. (1994) in an analysis of nearly 300 QSOs in the South Galactic Pole region (hereafter the SGP sample). While Hook et al. did not rule out the possibility of a positive correlation between variability and redshift in the SGP data, they concluded that the observed correlation was consistent with being entirely the result of the $L$–$z$ correlation propagated through the variability-luminosity relation. Though on balance the evidence seems to point towards a positive variability-redshift correlation, there is as yet no consensus (see discussion in Hook et al. 1994).

The relationship between variability and redshift is a key link between the variability properties of QSOs and of less luminous AGN. IUE observations of Seyfert 1 nuclei and low-redshift QSOs have demonstrated not only that variability is anticorrelated with luminosity, but also that the variations get larger towards shorter wavelengths (Edelson, Krolik & Pike 1990, Kinney et al. 1991, Paltani & Courvoisier 1994, Di Clemente et al. 1996). If this behaviour also holds for high-redshift QSOs, one would expect variability studies carried out at fixed observed wavelengths (say, the B-band) to show a positive correlation between variability and redshift (Giallongo et al. 1991, Hook et al. 1994, Trèvese et al. 1994, Cristiani et al. 1996, Di Clemente et al. 1996). As well as this connection with low-luminosity AGN, this effect would have important implications for the interpretation of the variability-luminosity relation: Neglecting the redshift/wavelength dependence would lead to an *overestimation* of the real rest-frame optical variability of high-$z$ objects, introducing an artificial flattening in the variability-luminosity relation, given the strong coupling between $L$ and $z$.

In this paper we analyse the variability-luminosity-redshift relationship for the QSOs in the SGP sample. Our approach differs from that of Hook et al. (1994) in that we study variability as a function of *both $L$ and $z$* using parametric fits to the data. The advantage of model-fitting over a correlation analysis is that the uncertainties in the measured quantities can easily be incorporated. Indeed, as we will demonstrate, photometric, sampling and k-correction uncertainties are a limiting factor in variability studies.

Our main goals in this study are:

- (i) to determine whether the data are consistent with a wavelength-dependent variability similar to that observed in low luminosity, nearby AGN, and
- (ii) to test Poissonian (or 'Christmas-tree') models for QSO variability. In its simplest version this model predicts a '$1/\sqrt{N}$' relationship between the rms variability and the mean luminosity. The proportionality constant carries information on the typical energy and time-scale of the individual pulses. More general models, in which the pulse properties scale with the QSO luminosity, are also investigated.

Section 2 introduces the variability indices used in this work. Section 3 presents a quantitative discussion of the variability-wavelength relation in nearby AGN, setting a reference point for the results of the fitting analysis. The data set employed is discussed in Section 4, while Section 5 presents the fitting method and a thorough discussion of the sources of error. The results of the analysis are presented in Section 6. The implications of our findings for Poissonian models are discussed in Section 7. Finally, in Section 8, we summarize our conclusions.

## 2 VARIABILITY INDICES

Two variability indices are used in this work. The first is

$$v \equiv \frac{\sigma(L_B)}{\overline{L_B}} \qquad (1)$$

where $\overline{L_B}$ is the mean B-band luminosity in the light curve and $\sigma(L_B)$ is its standard deviation.

The rms variability is not an ideal index for samples spanning a wide range of redshifts. Because of the time dilation factor $\Delta t_{rest-frame} = \Delta t_{obs}/(1+z)$, high-$z$ QSOs are observed for a substantially shorter rest-frame time than their low-$z$ counterparts. Since QSO variability is correlated within time scales smaller than $\sim$ 2–3 yr (Giallongo et al. 1991, Hook et al. 1994, Cristiani et al. 1996), time dilation alone could lead to a systematic decrease of $v$ for increasing $z$.

One way to avoid this bias is to use an index based on observations separated by a fixed rest-frame interval, or by an interval longer than a critical value (see also Giallongo et al. 1991, Cristiani et al. 1996). With this purpose we define our second variability index as

$$u \equiv \frac{\overline{[L_B(t_j) - L_B(t_i)]^2}^{1/2}}{\overline{L_B}} \qquad (2)$$

where $t_i$ and $t_j$ are epochs separated by more than 3 yr. This index corresponds to the asymptotic limit of the Structure Function (SF) of each individual QSO. The SFs of individual SGP objects are of course poorly defined given that only seven epochs are available. The *ensemble* SF, however, rises steadily for intervals shorter than $\approx$ 2–3 yr, becoming essentially flat (though fairly noisy) for longer time-scales (Hook et al. 1994, Cristiani et al. 1996, Giallongo et al. 1991, Trèvese et al. 1994).

## 3 VARIABILITY VERSUS WAVELENGTH IN NEARBY AGN

One of the goals of this paper is to compare the variability-wavelength relation of QSOs to that of low-luminosity, nearby AGN. In order to quantify this relationship we have parametrized it as a power law $v(\lambda) \propto \lambda^{-b}$ and used published data to estimate $b$ for nearby objects.

Among the five type-1 Seyferts studied by Edelson et al. (1990), four (NGC 4151, 3516, 5548 and Mrk 335) vary more at 1450 Å than at 2885 Å. We estimated $b$ for these AGN from

$$b = -\frac{\log{(v_{1450}/v_{2885})}}{\log{(1450/2885)}}$$

using $v_{1450}$ and $v_{2885}$ as listed in Table 1 of Edelson et al. (1990). The obtained values of $b$ range from 0.47 to 0.66, i.e., the variations are between 40 and 60 per cent larger at 1450 than at 2885 Å. Interestingly, 3C273 (the only luminous AGN in their sample) also shows a wavelength-dependent variability with $b \approx 1$.

Aretxaga, Cid Fernandes & Terlevich (1996) estimated $b$ from a comparison of the UV and optical light curves of NGC 4151 and 5548. They conclude that these nuclei vary 2–3 times more in the UV (1300–1800 Å) than in the B-band, which corresponds to $b$ between 0.6 and 1.

Finally, the results of Kinney et al. (1991) and Paltani & Courvoisier (1994) were also used to estimate $b$. This yielded $0.6 < b < 2.9$ (median $b = 1.3$) for Kinney et al.'s data (excluding blazars) and $0.2 < b < 1.7$ for Paltani & Courvoisier's data, but these values are less accurate than those quoted above.

The observations therefore point to values of $b$ between $\approx 0.5$ and 1.5 for nearby AGN. While a $v \propto \lambda^{-b}$ law is obviously a simple tentative parametrization, the derived values of $b$ provide an important reference point for the analysis of the $v(\lambda)$ relation in QSOs.

## 4 DATA: THE SGP SAMPLE

The SGP sample consists of 283 QSOs monitored since 1975 with the UK Schmidt telescope at the AAO (see Hook et al. 1994 for details of the observations). Redshifts are in the 0.43–4.01 range and apparent B magnitudes are between 17 and 21. Eleven plates were obtained, defining 7 different epochs spanning 16 yr. The rest-frame light curves are between 3.2 and 14.5 yr long allowing for time dilation. Alhough the sampling of each object is poor and the photometric errors are not ideal (typically $\pm 0.07$ mag), the large number of objects and wide coverage of the luminosity-redshift plane makes this sample one of the best QSO variability databases available.

The data consist of B-band light curves and redshifts for all 283 QSOs. Six objects were excluded from the original list, five of which had unreliable (or missing) $m_B$ values. The other object is 0055-2659, a highly variable luminous QSO, which seems to be a peculiar source (see discussion in Hook et al., who also excluded 0055-2659 from their analysis). For each of the remaining 277 QSOs we computed $\overline{L_B}$, $v$ and $u$. The k-correction was computed as in Hook et al., except for the fact that we shall investigate spectral indices other than $\alpha = -0.5$ ($F_\nu \propto \nu^\alpha$). $H_0 = 50$ km s$^{-1}$Mpc$^{-1}$ and $q_0 = 0.5$ are used throughout the paper, unless stated otherwise. Note, however, that $v$ and $u$ are *independent* of the k-correction and cosmological parameters, since they come in as multiplicative factors both in the numerator and denominator. Cosmology and k-correction affect only our estimate of $\overline{L_B}$, while $v$ and $u$ still measure the variability at $\lambda = \lambda_B/(1+z)$.

## 5 FITTING METHOD AND ERROR ANALYSIS

To examine the relationship between $v$, $\overline{L_B}$ and $z$ we have used parametric fits in the $(v, \overline{L_B}, z)$ space. This is done by fitting the data with a function of the form

$$v(\overline{L_B}, z) = A\overline{L_B}^a(1+z)^b \quad (3)$$

and analogously for $u(\overline{L_B}, z)$. The $(1+z)^b$ term is equivalent to $(\lambda/\lambda_B)^{-b}$, with $\lambda$ as the rest-frame wavelength in which the QSO is observed. The main advantage of equation (3), apart from its simplicity, is that its parameters can be used not only to test models for the $v(L)$ relation (Section 7), but also to compare the $v(\lambda)$ relation of QSOs to that observed in lower luminosity AGN (Section 3).

The fitting problem is linearized defining new variables:

$$\begin{cases} f \equiv \log v \\ x \equiv \log\left(\overline{L_B}/10^{12}L_B^{\odot}\right) \\ y \equiv \log(1+z) \end{cases} \quad (4)$$

Taking the logarithm of equation (3) we obtain

$$f(x,y) = ax + by + c \quad (5)$$

We therefore seek to fit a *plane* to the data points in the space $(x, y, f)$. The fit is performed using a $\chi^2$ minimization algorithm which takes into account the errors in both $f$ and $x$, but neglects errors in $y$. The $\chi^2$ reads as

$$\chi^2 = \sum_{i=1}^{N} \frac{(f_i - ax_i - by_i - c)^2}{\epsilon^2(f_i) + a^2\epsilon^2(x_i)}$$

Any meaningful model-fitting requires realistic estimates of the errors involved. Several sources of error affect each data point in the sample.

- $\overline{L_B}$: the k-correction necessary to bring all the observed luminosities to a common reference wavelength (the B-band in this case) introduces a substantial uncertainty in $\overline{L_B}$. The reason is that while a typical QSO spectrum needs to be assumed to compute the k-correction, individual QSOs have a wide range of spectral properties (e.g. Francis 1993). Reddening may also affect $\overline{L_B}$, although the indications are that QSOs suffer little intrinsic reddening (O'Brien, Gondhalekar & Wilson 1988, and references therein), and galactic reddening is likely to be low for the SGP objects.
- $v$ and $u$: the variability indices are affected by both the photometric errors and the sparse sampling of the light curves.
- $z$: the broad width of the emission lines in QSOs, as well as possible peculiar motions, introduce uncertainties in the measured redshifts. These, however, are much smaller than the ones in $v$ and $\overline{L_B}$, and we shall neglect them.

The uncertainties involved in deriving the mean luminosity of QSOs are not often allowed for when analysing the variability-luminosity relation. Recognizing the presence of such uncertainties has a strong impact on the resulting slopes. This is discussed by La Franca et al. (1995), who show that underestimating or neglecting the errors in an independent variable ($x$ in our case) leads to an *underestimation* of the true slope of the $f(x)$ relation, i.e. to a small absolute value of $a$.

In what follows we explain how we have estimated the uncertainties of $\overline{L_B}$ and $v$.

### 5.1 Errors in the luminosity

In the case of $\overline{L_B}$ we adopted the strategy of attributing the k-correction uncertainties to a *range of possible spectral indices.* Observed spectral indices in QSOs cover a wide range, from $\alpha \approx -2$ to $+2$, depending on which spectral window is used. The mean indices for the QSOs in the Large Bright QSO Survey (Morris et al. 1991, and references therein), for instance, are $\overline{\alpha} = -0.5$ for the 1455–2190 Å window and $-0.8$ for the 2190–3060 Å window, with an rms dispersion $\sigma_\alpha = 0.5$ for both wavelength regions (Francis 1993). In this study we shall assume that $\alpha$ follows a Gaussian distribution, centred at either $\overline{\alpha} = -0.5$ or $-1.0$, with $\sigma_\alpha = 0$, 0.25 or 0.5. For simplicity we follow the usual assumption that $\alpha$ is independent of both luminosity and redshift (but see Francis 1993).

As the k-correction increases with redshift, the uncertainty in $\alpha$ propagates to an uncertainty in $\overline{L_B}$ (and $x$) which increases with $z$. The k-correction used also allows for $Ly\alpha$ in emission as well as for absorption by intervening gas at $\lambda < 912$ Å (see Hook et al. 1994 for details). The uncertainty in $x$ was thus computed with a small simulation, evaluating $x$ for each QSO many times with $\alpha$ drawn from its Gaussian distribution. For $\sigma_\alpha = 0.25$ the typical error-bars in $x$ range from 0.08 to 0.17 for $1 \le z \le 4$, corresponding to errors of 18–39 per cent in $\overline{L_B}$.

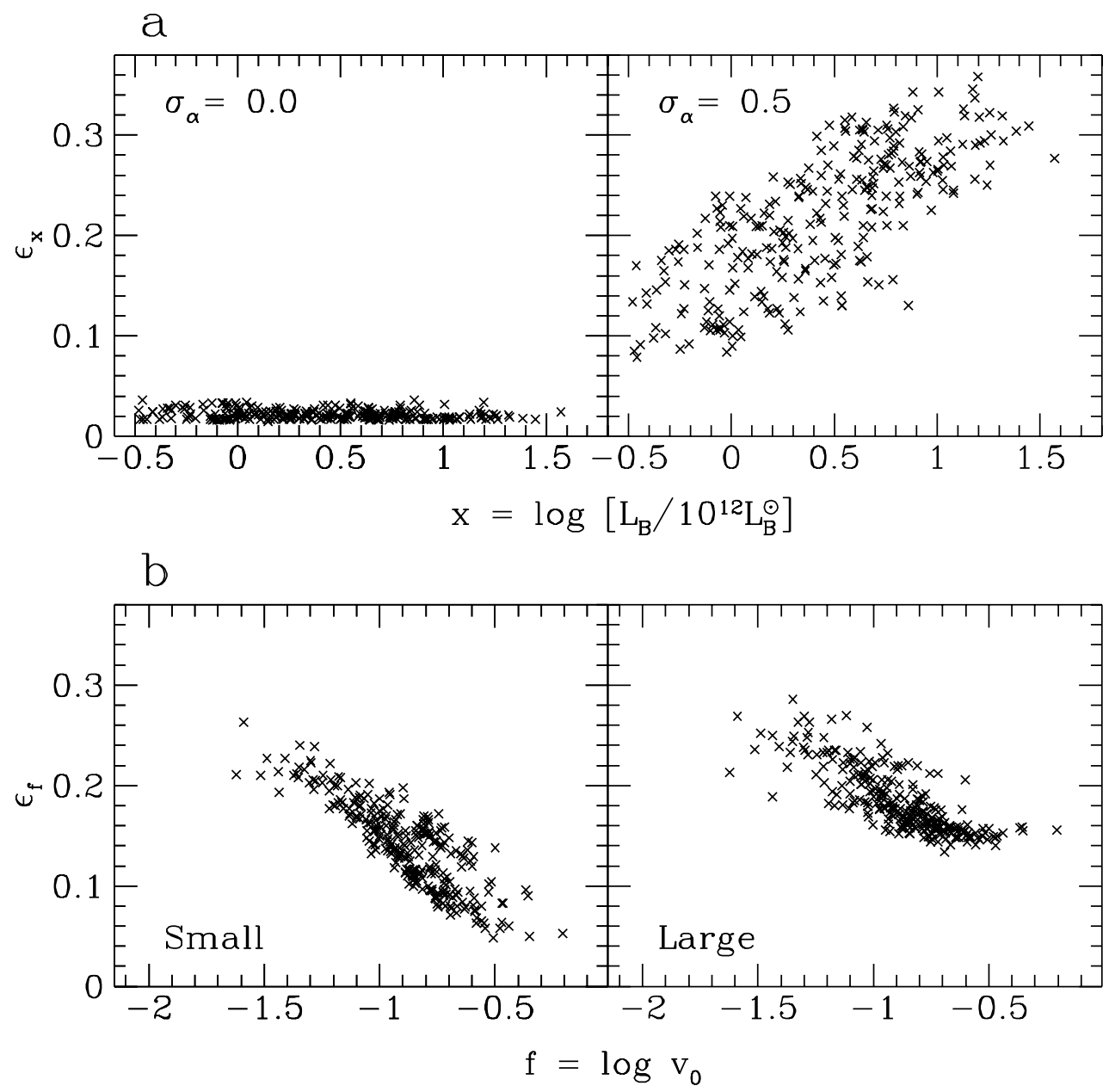

**Figure 1.** Extreme configurations for the errors in $x$ (a) and $\log v_0$ (b). The luminosities in panel (a) were computed with $\overline{\alpha} = -0.5$.

Each point in the light curve has a photometric error $\epsilon_m$ between 0.05 and 0.11 mag, depending on the object's apparent magnitude (see Table 2 in Hook et al. 1994). The mean luminosity obtained from the seven epochs is determined to an error $\sqrt{7}$ better than the individual points. This component, although it was small compared to the k-correction uncertainty, was included the evaluation of $\epsilon_x$ perturbing each point in the light curve with a $\Delta m_B$ drawn randomly from the true (apparent-magnitude dependent) error distribution of the SGP plates. The value of $x$ was then computed 1000 times for different realizations of the perturbations and spectral indices, yielding a reliable estimate of the combined photometric + k-correction uncertainties upon $x$.

### 5.2 Errors in the variability index

The photometric noise contribution to the rms variability was subtracted in quadrature from the measured rms: $v_0 = (v^2 - \epsilon_v^2)^{1/2}$, with $\epsilon_v = 2.5 \log e \times \epsilon_m$. Six QSOs (all of which are in the high luminosity half of the sample) were excluded from the analysis because they varied *less* than their estimated photometric error.

The quadrature correction only improves our estimate of $v$; it does not provide a measure of how accurate this estimate is. The uncertainty in $v_0$ is most easily evaluated with a Monte Carlo simulation. For each object we: (1) perturb each point in the light curve by an error drawn from the appropriate error distribution, (2) compute $v$ and $v_0$, and (3) repeat the above steps 1000 times. The standard deviation of the resulting $v_0$'s gives a realistic estimate of $\epsilon_{v0}$.

The error estimates obtained above are entirely due to photometric noise. The poor sampling of the light curves ought to introduce a further uncertainty in $v_0$. An estimate of the total photometric + sampling uncertainties upon $v_0$ can be obtained assuming that QSO variability is a Gaussian process, i.e. that $L_B(t)$ is a Gaussian-random variable with mean $\overline{L_B}$ and standard deviation $\sigma_0 = v_0 \times \overline{L_B}$. The procedure here was, for each object, to: (1) compute $\overline{L_B}$, $v$ and $v_0$, (2) simulate a 7 points Gaussian light curve $L_B(t)$ with standard deviation $\sigma_0$, (3) add a perturbation drawn from the appropriate error distribution and recompute $v_0$ and (4) repeat the above steps 1000 times, obtaining a new estimate for $\epsilon_{v0}$. This new estimate measures the degree to which we can measure the correct standard deviation of a Gaussian process given only 7 points, all of which are affected by photometric errors. This method provides an upper limit to $\epsilon_{v0}$, since QSO variability is *not* Gaussian in the time scales sampled by the SGP data, otherwise the ensemble SF should be flat, showing no correlated variability at all. On the contrary, the SF rises during the first 2–3 yr and only then flattens to a roughly constant level (Hook et al. 1994). Since many of the observations were carried out within (rest-frame) intervals smaller than 3 yr, we expect the observed fluctuations to be smaller than the ones corresponding to a truly Gaussian process.

The values of $\epsilon_{v0}$ obtained with the 'Gaussian-method' are typically $\sim$ 40 per cent larger than the ones obtained considering the photometric errors alone. The true uncertainties in $\epsilon_{v0}$ lie between these two estimates. The errors in $f = \log v_0$ are easily evaluated either by using a standard propagation of errors or by directly computing $f$ in the simulations.

## 6 RESULTS

It should be clear from the discussion in Section 5.1 and Section 5.2 that though the sources of errors in luminosity ($x$) and variability index ($f$) are well understood, the exact magnitude of these errors is not well defined. It is therefore important to investigate how sensitive the best fitting parameters $a$, $b$ and $c$ are to the choice of errors. We shall discuss nine different combinations of the errors in $x$ and $f$:

- $\epsilon_x$: $\sigma_\alpha = 0,\ 0.25,\ 0.5$
- $\epsilon_f$: Small, Large, Intermediate

where 'Small', 'Large' and 'Intermediate' correspond to errors $\epsilon_{v0}$ computed using only the photometric errors, photometric + sampling uncertainties, and the mid-point between these two extremes, respectively. Fig. 1 illustrates some of these configurations. Less variable objects (low $f$) have systematically larger $\epsilon_f$'s due to the increasingly uncertain quadrature correction to the variability indices. More luminous objects have larger $\epsilon_x$, since they have higher redshifts and are thus more affected by the k-correction uncertainties. In addition, two values for the mean spectral index $\overline{\alpha}$ will be investigated: $-0.5$, the value used by Hook et al. (1994), and $-1$, the value used in Trèvese et al. (1994).

The data is shown in Fig. 2, where we plot $f$ versus $y$, $f$ versus $x$ and $y$ versus $x$. The left and right columns show the measurements with the smallest and largest possible errors allowed in this study, respectively. Fig. 3 attempts a visualization of the variability in the luminosity-redshift plane. Each QSO is represented by a square whose size scales with $v_0$. The anticorrelation between variability and luminosity is clearly visible: squares are smaller towards larger $x$. The variability-redshift relation shows a large scatter, though the tendency is for $v_0$ to increase with $y$ within fixed luminosity bands.

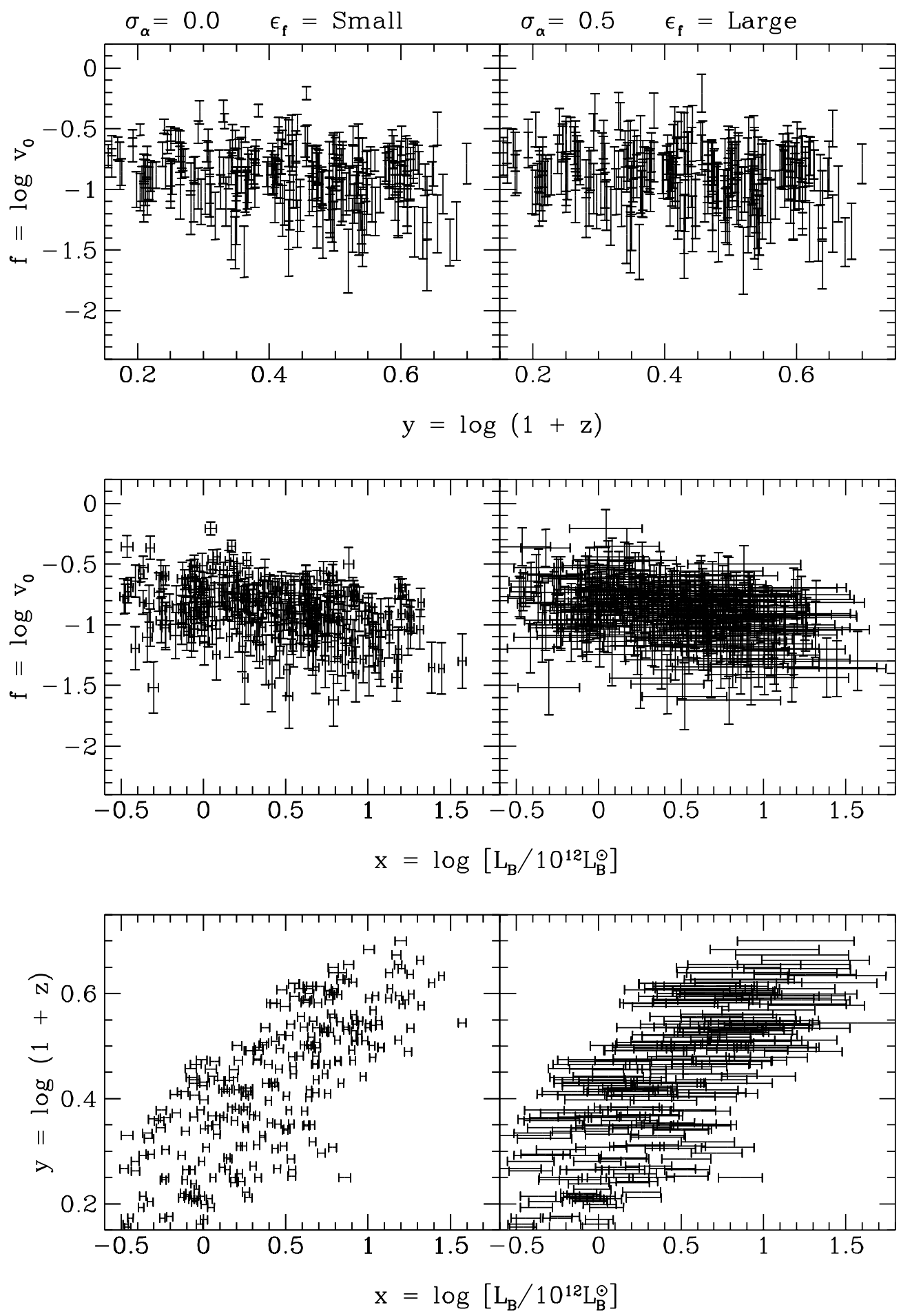

**Figure 2.** $\log v_0 \times y$, $\log v_0 \times x$ and $x \times y$ for the QSOs in the SGP sample. Error bars correspond to $(\sigma_\alpha, \epsilon_f) = (0, Small)$ for the plots in the left and $(0.5, Large)$ for those in the right.

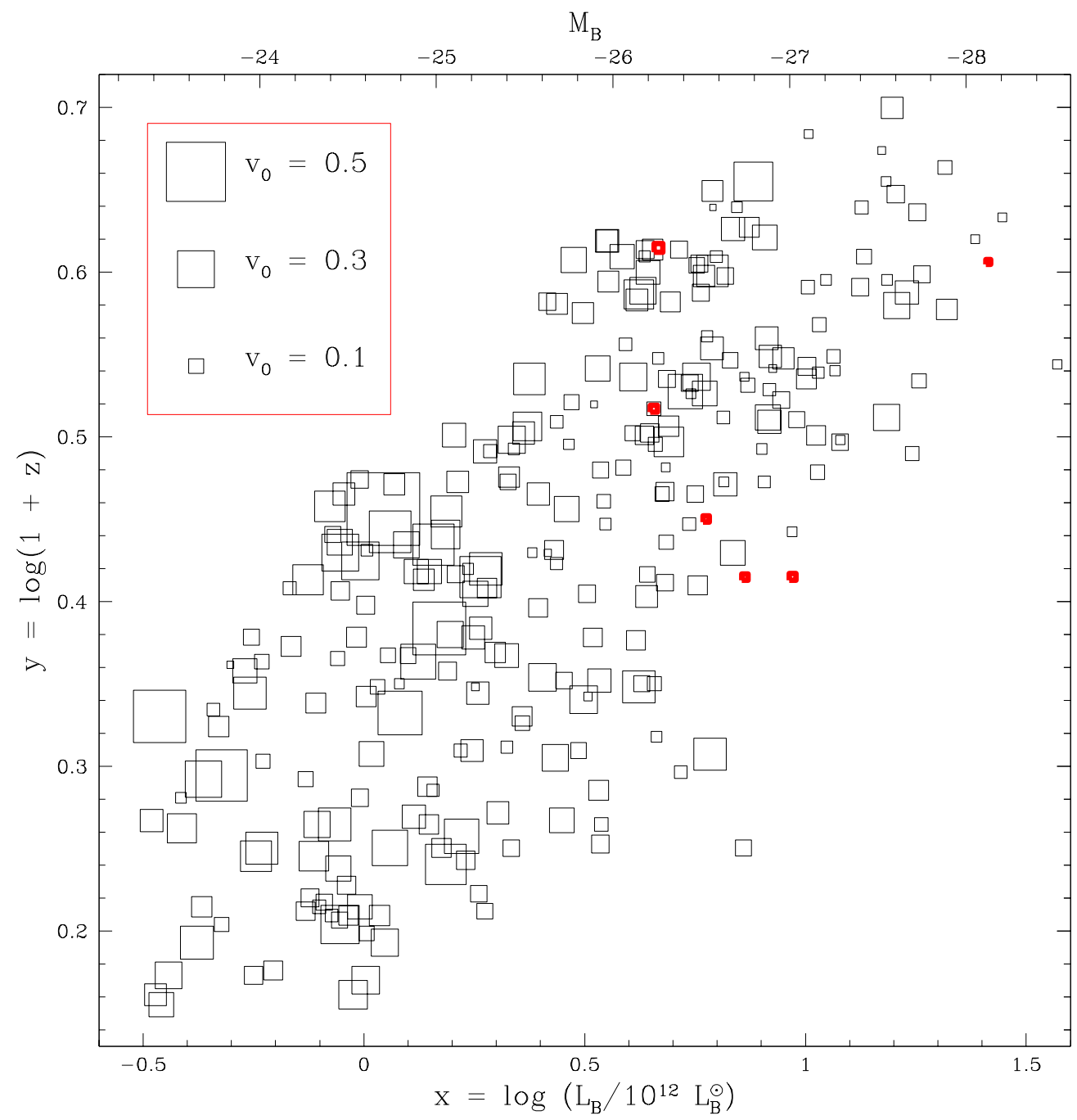

**Figure 3.** Projected representation of the variability in the $x \times y$ plane. The size of each square scales linearly with $v_0$. The six objects marked with filled squares are the ones which vary less than the expected photometric error, with $v_0$ represented by $|v^2 - \epsilon_v^2|^{1/2}$. The absolute magnitude corresponding to $\overline{L_B}$ is shown in the upper axis.

The results of the parametric fits are presented in Table 1. *The best fitting parameters a, b and c are very sensitive to the choice of errors.* In general $a$ gets increasingly negative and $b$ increases as $\sigma_\alpha$ (and consequently $\epsilon_x$) increases, whereas the opposite happens as $\epsilon_f$ increases. The parameters $a$ and $b$ always vary in opposite directions due to the strong coupling between luminosity and redshift. The independent term $c$ tends to follow $a$ in its dependence on $\epsilon_x$ and $\epsilon_f$. While $a$ and $c$ are nearly unaffected by the mean spectral index used in the k-correction, $b$ increases as $\overline{\alpha}$ goes from $-0.5$ to $-1$.

The changes in the best-fitting parameters resulting from different choices of errors are substantial. For $\overline{\alpha} = -0.5$, for instance, the pair $(a, b)$ changes from $(-0.81, 1.52)$ to $(-0.19, 0.12)$ as we move from the case in which $\epsilon_x$ is large and $\epsilon_f$ is small to the opposite extreme. The errors in the fit parameters (see Fig. 4) are smaller than this range. It is therefore clear that the uncertainty in $a$, $b$ and $c$ is dominated by our lack of detailed knowledge of the magnitude of the errors involved. On the assumption that Table 1 covers the whole spectrum of possibilities for the errors, and discarding cases where either $f$ or $x$ carries essentially all the errors (i.e., maximum $\epsilon_f$ and minimum $\epsilon_x$ and vice-versa), we can safely bracket the best-fitting parameters within the following intervals:

$$\left\{ \begin{array}{ccc} -0.6 & < a < & -0.2 \\ 0.1(0.2) & < b < & 1.0(1.3) \\ -1.0 & < c < & -0.7 \end{array} \right. \quad (6)$$

where values of $b$ outside parentheses correspond to $\overline{\alpha} = -0.5$ fits, and those inside are for $\overline{\alpha} = -1$. Unless we choose a specific error-configuration, it is not possible to determine $a$, $b$ and $c$ to a better degree of precision.

Despite the large ranges in the fitting parameters, two strong conclusions can be derived from our analysis. First, we confirm that variability and luminosity are anticorrelated in the SGP QSOs ($a < 0$ in all fits), as previously found by Hook et al. (1994—see also Pica & Smith 1983, Cristiani, Vio & Andreani 1991, Cristiani et al. 1996). Secondly, the SGP data point to the existence of a positive correlation between variability and redshift ($b > 0$ in all fits), in agreement with Cristiani et al. (1996), who used a different method of analysis. Furthermore, the obtained values of $b$ are consistent with the ones estimated from the variability-wavelength relation observed in less luminous AGN ($0.5 < b < 1.5$). This is a strong indication that the variability-wavelength anticorrelation observed in nearby AGN is also followed by high-redshift QSOs (see also Paltani & Courvoisier 1994, Di Clemente et al. 1996).

The uncertainties in $a$, $b$ and $c$ are illustrated in Fig. 4, where we plot the one, two and three sigma confidence levels in the $a$ versus $b$ and $a$ versus $c$ planes for the four extreme combinations of $\sigma_\alpha$ and $\epsilon_f$: $(\sigma_\alpha, \epsilon_f) = (0, Small)$, $(0.5, Small)$, $(0, Large)$ and $(0.5, Large)$. Other combinations of the errors in $x$ and $f$ yield contours within the general area delimited by these cases. The goodness of the fits, measured by the reduced $\chi^2$, are listed in Table 1. An acceptable $\chi^2$ here means $\chi^2/d.o.f < 1.288$ given the 268 degrees of freedom—the probability of larger values occurring by chance is $10^{-3}$. Only two of the fits produce acceptable $\chi^2$ values, notably (but not surprisingly) those with large $\epsilon_x$ and $\epsilon_f$. In any case, though the uncertainties in $a$, $b$ and $c$ are large, it is clear that $a > 0$ and $b < 0$ models are ruled out by the observations.

| $\overline{\alpha}$ | $\sigma_\alpha$ | $\epsilon_f$ | $a$ | $b$ | $c$ | $\chi^2$/d.o.f. |
|---|---|---|---|---|---|---|
| -0.5 | 0.00 | S | -0.24 | 0.20 | -0.74 | 2.998 |
| -0.5 | 0.00 | I | -0.20 | 0.13 | -0.78 | 1.832 |
| -0.5 | 0.00 | L | -0.19 | 0.12 | -0.81 | 1.351 |
| -0.5 | 0.25 | S | -0.40 | 0.57 | -0.85 | 2.661 |
| -0.5 | 0.25 | I | -0.28 | 0.32 | -0.83 | 1.780 |
| -0.5 | 0.25 | L | -0.24 | 0.24 | -0.84 | 1.332 |
| -0.5 | 0.50 | S | -0.81 | 1.52 | -1.15 | 1.765 |
| -0.5 | 0.50 | I | -0.59 | 1.02 | -1.03 | 1.446 |
| -0.5 | 0.50 | L | -0.46 | 0.76 | -0.98 | 1.180 |
| -1.0 | 0.00 | S | -0.24 | 0.28 | -0.73 | 2.992 |
| -1.0 | 0.00 | I | -0.20 | 0.22 | -0.77 | 1.833 |
| -1.0 | 0.00 | L | -0.20 | 0.24 | -0.81 | 1.351 |
| -1.0 | 0.25 | S | -0.41 | 0.79 | -0.86 | 2.668 |
| -1.0 | 0.25 | I | -0.28 | 0.46 | -0.83 | 1.790 |
| -1.0 | 0.25 | L | -0.24 | 0.35 | -0.84 | 1.337 |
| -1.0 | 0.50 | S | -0.78 | 1.87 | -1.14 | 1.751 |
| -1.0 | 0.50 | I | -0.59 | 1.32 | -1.03 | 1.464 |
| -1.0 | 0.50 | L | -0.46 | 1.00 | -0.98 | 1.176 |

**Table 1.** Best fitting parameters $a$, $b$ and $c$ and reduced $\chi^2$ for the $\log v_0 = ax + by + c$ fits. The different lines correspond to different combinations of $\overline{\alpha}$, $\sigma_\alpha$ and $\epsilon_f$. S, I and L denote small, intermediate and large errors in $f = \log v_0$.

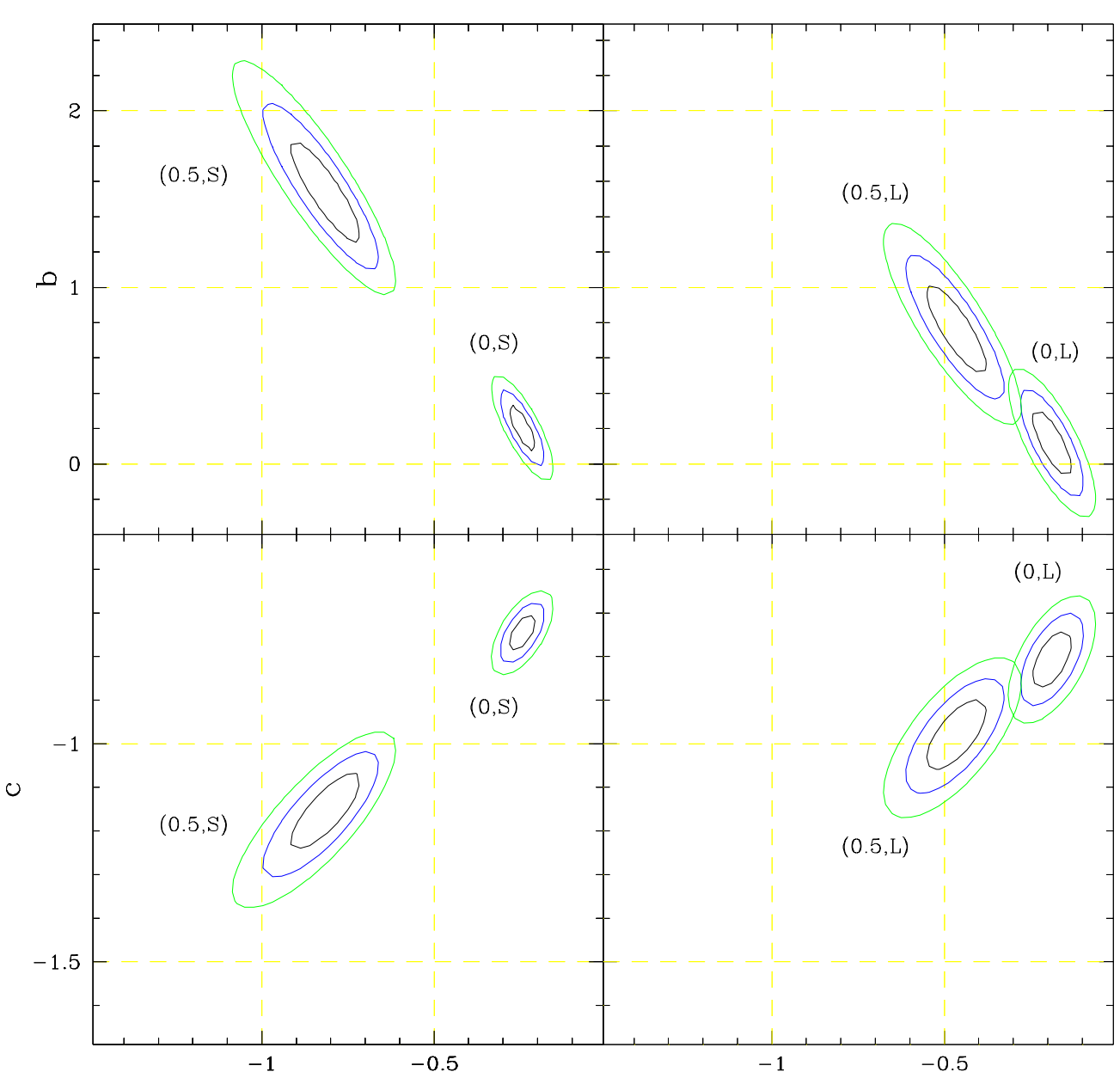

**Figure 4.** $\Delta\chi^2$ maps for the $v(\overline{L_B}, z)$ fitting parameters, showing the 68.3, 95.4 and 99.7 per cent confidence level contours. Numbers in parentheses are $(\sigma_\alpha, \epsilon_f)$. All contours are for $\overline{\alpha} = -0.5$.

### 6.1 Tests of the fits

A few variations over the fitting analysis were tried to test the sensitivity of the fits to some aspects of the data.

The k-correction includes a correction for absorption by the $Ly\alpha/Ly\beta$ forest and for Lyman-limit systems (Irwin, McMahon & Hazard 1991). Since this correction affects only the QSOs with $z > 3.28$, we have experimented with excluding the 17 such objects from the sample to test how strongly they affect the fits. The resulting changes in $a$, $b$ and $c$ were negligible, typically less than 0.1.

Fits were also made to subsets of the data dividing the sample into luminosity and redshift halves. The values of $a$ for the lower luminosity and redshift half-samples were systematically less negative than the values for the high luminosity and redshift halves. As usual, $b$ showed the opposite behaviour. Whilst this result indicates a *bending* of the $f(x, y)$ plane, the major factor affecting the fits is the increase in the errors as we go towards more luminous, high redshift objects. In fact, the low-$L$ and low-$z$ fits are closer to the ones obtained for the full sample, since these objects weight more in the fits as they have smaller errors. A wider coverage of the luminosity-redshift plane and smaller errors would be required to properly assess the reality of this bending.

We also computed fits neglecting the presence of errors in $x$ and $f$. These yielded $a = -0.26$ and $c = -0.87$, whereas $b$ was 0.19 for $\alpha = -0.5$ and 0.31 for $\alpha = -1$. Finally, fits were also performed for $H_0 = 100$ km s$^{-1}$Mpc$^{-1}$. As expected, the coefficients $a$ and $b$ are not affected by $H_0$, but $c$ is systematically smaller by $\Delta c = a \log 4$ than for $H_0 = 50$ km s$^{-1}$Mpc$^{-1}$. This difference amounts to $\approx$ 0.1–0.5 for the values of $a$ in Table 1.

### 6.2 Fits with a fixed wavelength/redshift dependence

Instead of allowing for a wavelength dependence in the fits, one could alternatively start with the assumption that QSOs follow the same variability-wavelength relationship observed in nearby AGN. Adopting a fixed value of $b$, we can apply a *correction* to the observed variability indices to convert them to a common wavelength for all QSOs and then examine the consequences for the variability-luminosity relationship.

In Fig. 5 we plot the best-fitting values of $a$ and $c$ as a function of $b$ for different combinations of $\sigma_\alpha$ and $\epsilon_f$. The tendency is for both $a$ and $c$ to decrease with increasing $b$. Adopting a value of $b$ similar to those found in nearby AGN $(0.5 < b < 1.5)$ would result in $-0.8 < a < -0.3$, while $c$ would lie in the $-1.2$ to $-0.8$ interval. The $\Delta\chi^2$ profiles are also shown in Fig. 5 to illustrate the change in the quality of the fit for different values of $b$.

Note that even for $b = 0$, i.e., no wavelength dependence, the slope $a$ is critically dependent on the choice of errors, particular for the luminosity. The change in $a$ between fits with $\sigma_\alpha = 0$ and 0.5 is comparable to that resulting from changing $b$ from 0 to 1.

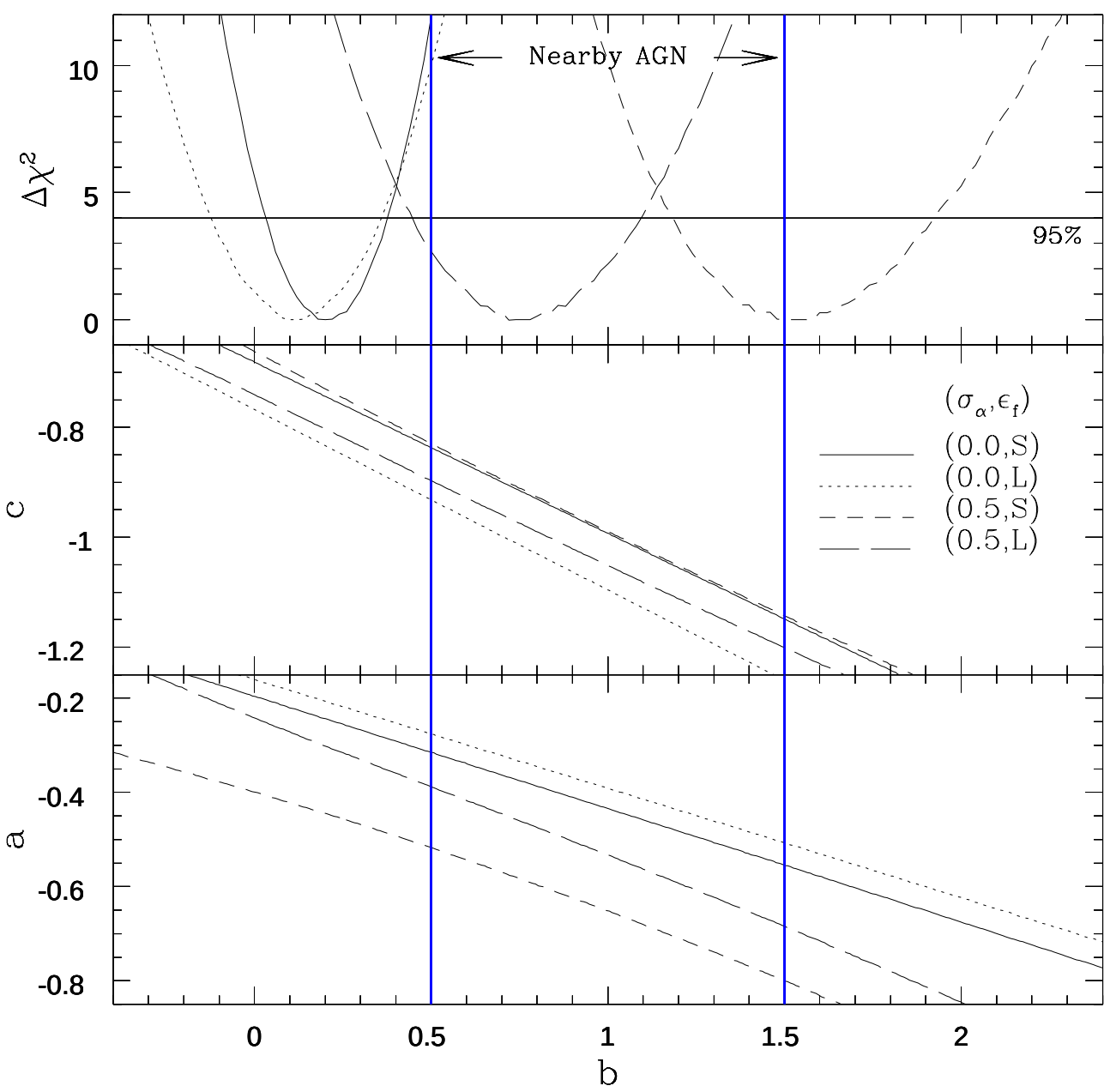

**Figure 5.** Best-fitting parameters $a$ and $c$ as a function of $b$ for fixed $b$ fits for different combinations of the errors in luminosity ($x$) and variability index ($f$). The top panel shows the change in the $\chi^2$ with respect to the model with minimum $\chi^2$. The range of values of $b$ found in nearby AGN is indicated for comparison.

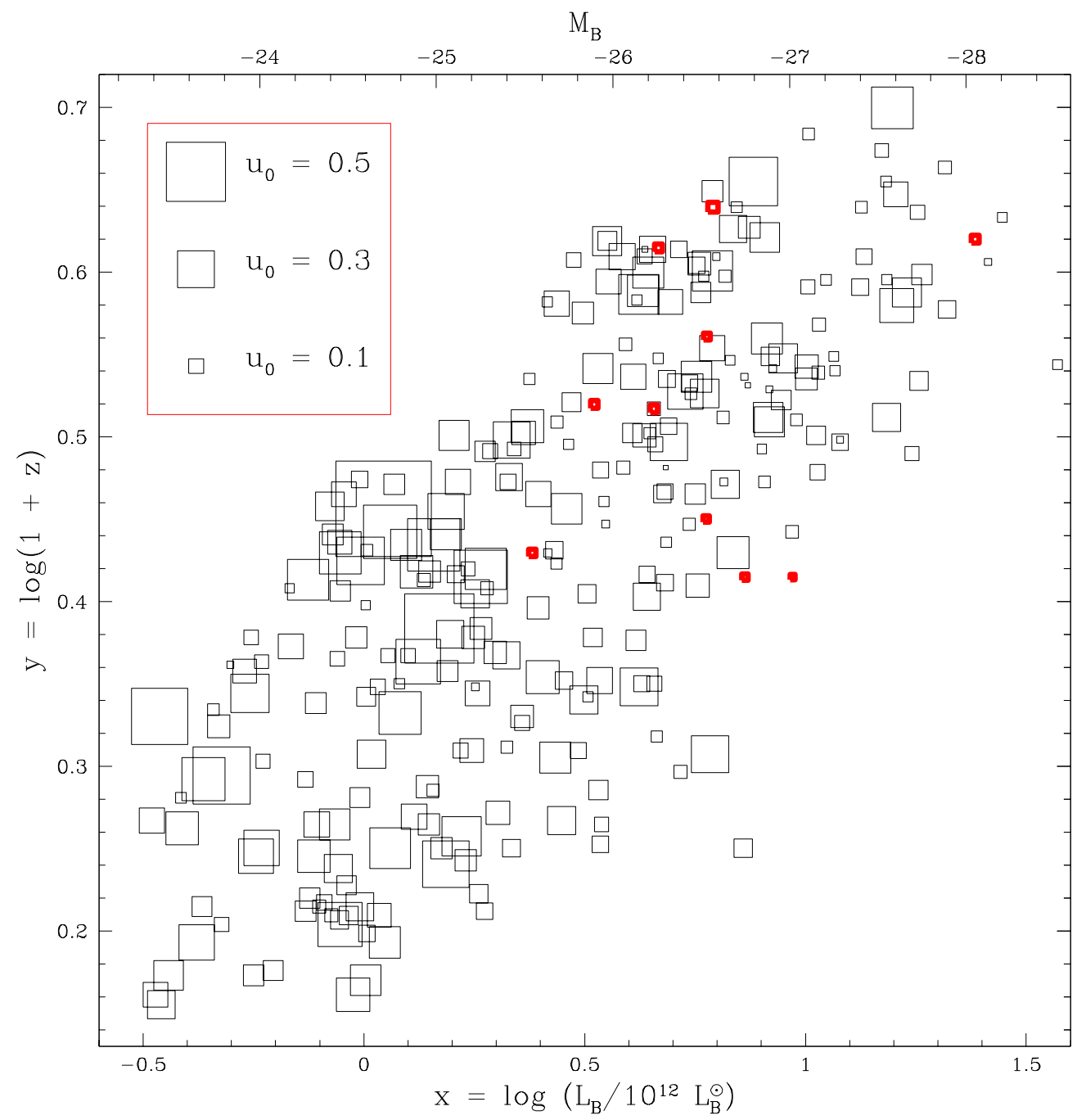

**Figure 6.** As Fig. 3, but for variability index $u_o$.

| $\overline{\alpha}$ | $\sigma_\alpha$ | $\epsilon_f$ | $a$ | $b$ | $c$ | $\chi^2$/d.o.f. |
|---|---|---|---|---|---|---|
| -0.5 | 0.00 | S | -0.26 | 0.27 | -0.71 | 3.980 |
| -0.5 | 0.00 | I | -0.19 | 0.08 | -0.73 | 2.375 |
| -0.5 | 0.00 | L | -0.18 | 0.04 | -0.75 | 1.771 |
| -0.5 | 0.25 | S | -0.48 | 0.77 | -0.86 | 3.430 |
| -0.5 | 0.25 | I | -0.30 | 0.33 | -0.79 | 2.320 |
| -0.5 | 0.25 | L | -0.25 | 0.19 | -0.78 | 1.749 |
| -0.5 | 0.50 | S | -1.13 | 2.23 | -1.30 | 2.077 |
| -0.5 | 0.50 | I | -0.81 | 1.48 | -1.11 | 1.814 |
| -0.5 | 0.50 | L | -0.61 | 1.02 | -1.01 | 1.528 |
| -1.0 | 0.00 | S | -0.25 | 0.36 | -0.70 | 3.946 |
| -1.0 | 0.00 | I | -0.20 | 0.17 | -0.72 | 2.385 |
| -1.0 | 0.00 | L | -0.19 | 0.17 | -0.75 | 1.769 |
| -1.0 | 0.25 | S | -0.50 | 1.04 | -0.86 | 3.447 |
| -1.0 | 0.25 | I | -0.30 | 0.48 | -0.79 | 2.340 |
| -1.0 | 0.25 | L | -0.24 | 0.32 | -0.79 | 1.741 |
| -1.0 | 0.50 | S | -1.12 | 2.76 | -1.29 | 2.077 |
| -1.0 | 0.50 | I | -0.82 | 1.92 | -1.12 | 1.830 |
| -1.0 | 0.50 | L | -0.61 | 1.33 | -1.01 | 1.523 |

**Table 2.** As Table 1 but for the $\log u_0 = ax + by + c$ fits.

### 6.3 Results for the SF-based variability index

Fits analogous to those for $v$ were also performed for the SF-based variability index, $u$, defined in equation (2). The analysis of the errors in $u$ was carried out as for $v$, with the exception that the quadrature correction to $u$ is $u_0 = (u^2 - \epsilon_u^2)^{1/2}$, with $\epsilon_u = 5 \log e \times \epsilon_m$. This time, 10 objects were excluded from the analysis for having $u < \epsilon_u$. These are marked with filled squares in Fig. 6, where, analogously to Fig. 3, we represent the variability in the $x \times y$ plane with squares whose sizes scale with $u_0$. We find that $\epsilon_{u0}$ is typically 10–20 per cent larger than $\epsilon_{v0}$, and that, as for $v_0$ (Fig. 1), objects with small $u_0$ are subjected to larger errors. Fig. 7 shows the variability versus luminosity and redshift relations for $u_0$ in the cases of minimum and maximum errors.

The results of the fits are listed in Table 2, and confidence contours are plotted in Fig. 8. It is noticeable that $a$ tends to be more negative and $b$ more positive than for the $v_0$-fits. This happens essentially because $\log u_0$ covers a larger dynamic range than does $\log v_0$, which naturally implies larger slopes. The range of $\log u_0$ in the sample is from $-2.1$ to $-0.1$, with an rms dispersion of 0.3, whereas for $\log v_0$ the range is $-1.6$ to $-0.2$ and the rms is 0.2. Notice, however, that this difference is only noticeable for fits with large $\sigma_\alpha$. Fits with no errors resulted in $a = -0.24$ and $c = -0.80$, with $b = 0.05$ for $\overline{\alpha} = -0.5$ and 0.17 for $\overline{\alpha} = -1$.

Fig. 9 shows the ratio between the two variability indices as a function of $v_0$, $z$ and $x$. The index $u_0$ is typically $\sim 20$ per cent larger than $v_0$ since it is based on variations over a longer time base-line, but there is no obvious trend of the ratio $u_0/v_0$ with $v_0$, $z$ or $x$. This indicates that $v_0$ is not severely affected by the time-dilation bias discussed in Section 2, otherwise $u_0/v_0$ should exhibit a systematic increase with $z$. This is not seen in Fig. 9, with the possible exception of $z > 3$ objects. This, however, is not necessarily a general result, as it could be the result of the particular distribution of time intervals in the SGP sample. The scatter in $u_0/v_0$ increases substantially for luminous, high-$z$, less variable objects, most probably due to the larger errors, particularly in $u_0$. Because of time-dilation, the number of pairs of plates involved in the computation of $u_0$ decreases from 15 to 7 as $z$ increases from 0.4 to 3.6, and for the highest redshift object in the sample $u_0$ is computed from only two epochs.

Analogously to equation (6), we can bracket the best-

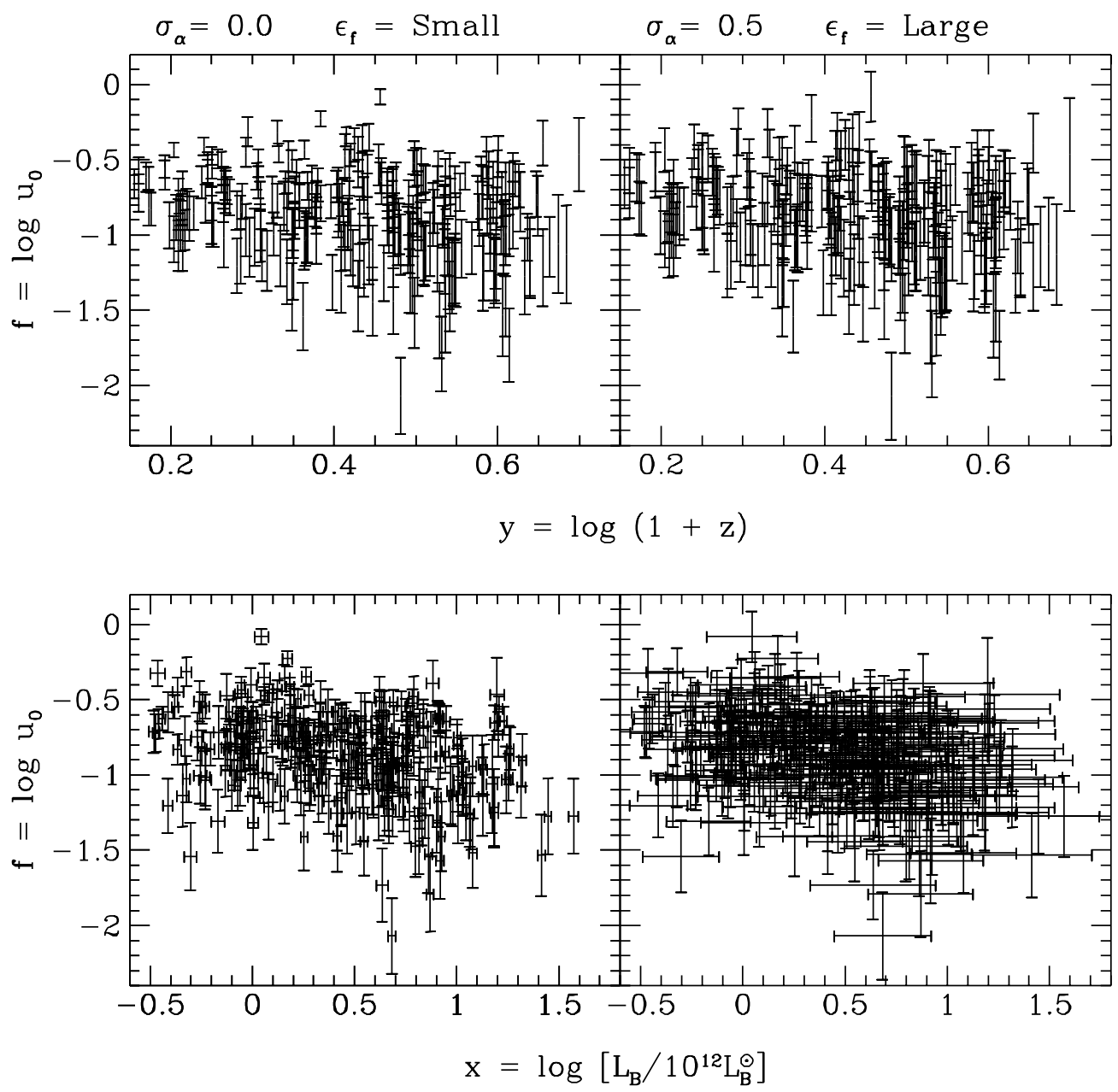

**Figure 7.** As Fig. 2 but for $u_0$ as the variability index.

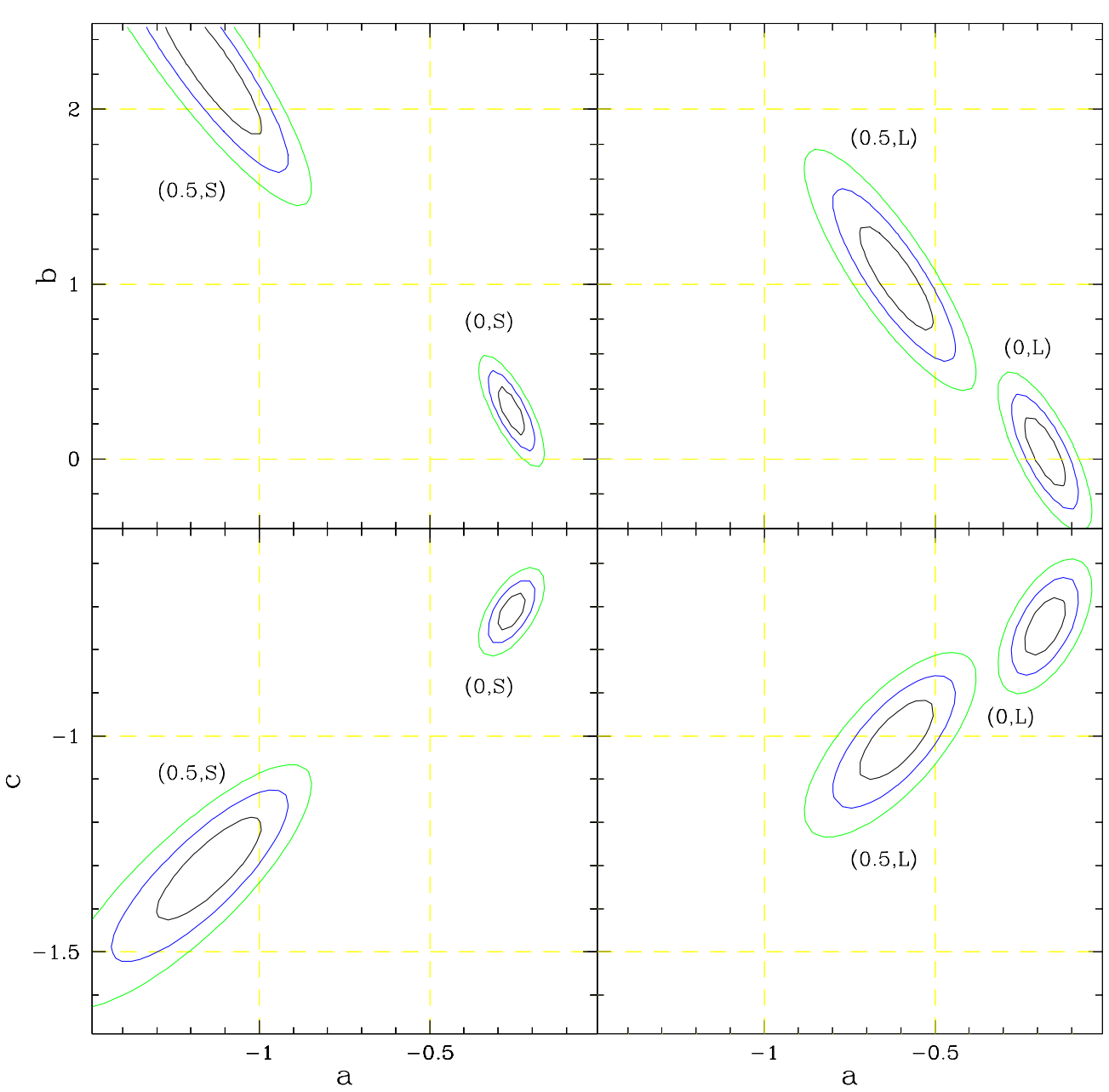

**Figure 8.** As Fig. 4, but for the $u_0(\overline{L_B}, z)$ fits.

fitting parameters for the $u_0$ fits between the following intervals:

$$\left\{ \begin{array}{ccc} -0.8 & < a < & -0.2 \\ 0.1(0.2) & < b < & 1.5(1.9) \\ -1.1 & < c < & -0.7 \end{array} \right. \quad (7)$$

(where values of $b$ outside parentheses correspond to $\overline{\alpha} = -0.5$ fits and those inside are for $\overline{\alpha} = -1.0$). Note, however, that the $\chi^2$ values in Table 2 are larger than those for the $v_0$-fits, which indicates that our fitting equation is too simplistic and/or other quantities besides $L$ and $z$ are involved in defining the variability properties of QSOs.

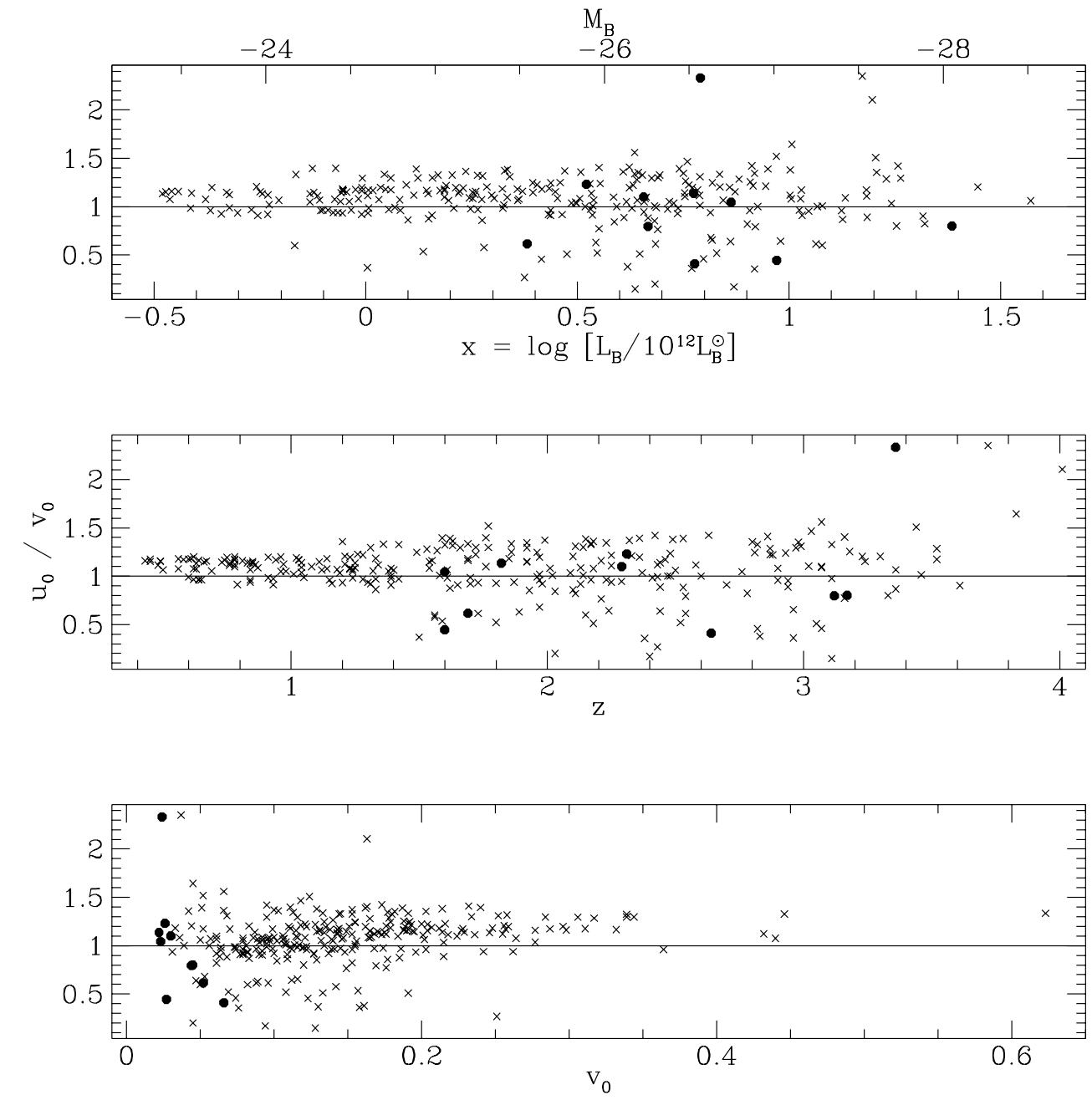

**Figure 9.** Ratio of the two variability indices, $u_0$ and $v_0$ as a function of $v_0$, redshift and luminosity. Filled circles indicate objects for which $v < \epsilon_v$ and/or $u < \epsilon_u$. The upper scale in the top panel shows the absolute magnitude corresponding to $x$.

## 7 TESTING MODELS FOR QSO VARIABILITY

So far we have focused on the analysis of the data. In what follows we shall demonstrate that, despite the uncertainties discussed, the SGP sample imposes interesting constraints on theoretical models for AGN variability.

Testing the consistency of Poissonian models for the optical variability of QSOs is of major interest in this study. In this general class of models variability is regarded as beibg the result of the random superposition of uncorrelated pulses. In this section we use the results of our fits to test the validity of this scenario and to constrain its basic parameters.

The essential elements of a Poissonian model are the rate, energy and time-scale of the pulses. These are denoted by $\nu$, $E_B^p$ and $\tau_B^p$ respectively, the subindex indicating that we are referring to the B-band. The hypothesis that the events are randomly distributed in time yields the following expressions for the mean B-band luminosity and net rms deviation:

$$\overline{L_B} = \nu \overline{E_B^p} / f_{var} \quad (8)$$

and

$$v_B = f_{var}^{1/2} \left( \overline{L_B} / \overline{L_B^p} \right)^{-1/2}, \quad (9)$$

where $\overline{L_B^p} \equiv \overline{E_B^p}/\overline{\tau_B^p}$ denotes the typical luminosity of an individual pulse, with $\overline{E_B^p}$ and $\overline{\tau_B^p}$ as the mean B-band energy and lifetime of the pulses respectively (Cid Fernandes 1995). Equations (8) and (9) allow for the possibility that a fraction $f_{bck} = 1 - f_{var}$ of the mean luminosity $\overline{L_B}$ is due to a constant 'background' component in the light curve. It is clear that equation (9) is simply a 'one over square root of $N$' law, diluted by the fact that a fraction $f_{bck}$ of the observed luminosity is not variable. In what follows we use the SGP data to constrain $f_{bck}$, $\overline{E_B^p}$ and $\nu$.

### 7.1 The background fraction: $f_{bck}$

Several sources can act as a constant background in AGN light curves: starlight from the host galaxy, a nuclear/circumnuclear star cluster, scattering, and/or the non-variable part of the nuclear engine. Starlight from the host galaxy is certainly present in the light curves of QSOs, but its contribution to the total luminosity is likely to be small, as indicated both by direct imaging studies (Aretxaga, Boyle & Terlevich 1995, Hutchings 1995) and by the absence of strong stellar features in the spectra of QSOs. The nuclear engine itself, however, could well have a non-variable component. This is, for instance, the case in the starburst model for AGN; in this model a substantial fraction of the optical-UV luminosity is the result of the presence of young stars (Aretxaga & Terlevich 1994). In an accretion disc model, one can also expect that part of the disc will remain unaffected by the instabilities promoting the variability, thus playing the role of a background component. Scattering of the nuclear photons in an extended region, a central ingredient in current unified models (e.g. Antonucci 1993), would smooth out the variations and would also appear as a constant background.

An upper limit for the background fraction can be obtained directly from the data using the average and minimum luminosities in the light curves. Since $L_B^{min} \geq L_B^{bck}$ and $L_B^{bck} = f_{bck}\overline{L_B}$, we have that

$$f_{bck} \leq \frac{L_B^{min}}{\overline{L_B}}$$

Fig. 10 shows the results of this exercise. As expected, $L_B^{min}/\overline{L_B}$ increases with luminosity, since the chances of a light curve reaching its true minimum decrease as the rate of events increases. Accordingly, the less luminous objects are those with a smaller upper limit for $f_{bck}$. Taking this systematic effect into account, we estimate that $f_{bck} < 0.7$ for all QSOs. Note that several points in Fig. 10 lie above this limit. This conservative estimate allows for the possibility that $L_B^{min}$, which is evaluated from a single point in the light curve, may be affected by the photometric errors. This is an important *model-independent* quantity for variability studies. If QSOs do contain a constant light source aside from the variable component, then this non-variable source contributes at most 70 per cent of their B-band luminosity. As a comparison, the ratio of minimum to average B luminosities in NGC 4151 and 5548 is $\approx 0.5$ in both cases, implying $f_{bck} < 0.5$ (Aretxaga & Terlevich 1993, 1994, Cid Fernandes, Terlevich & Aretxaga 1996). Given the lower rate of events in these objects, this estimate is probably closer to its true value for them than it is for QSOs.

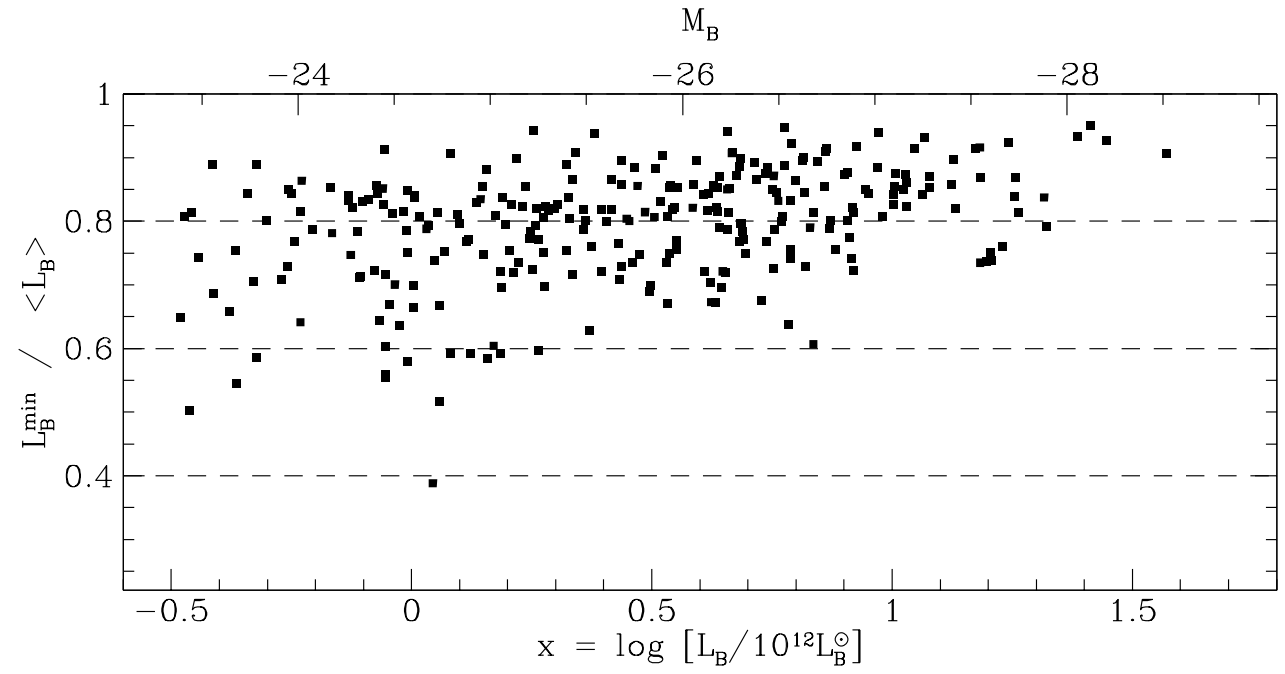

**Figure 10.** Upper limits for the background fraction ($f_{bck}$) for the QSOs in the SGP sample.

### 7.2 The Simple Poissonian model: Constraints on the energy of the pulses

The simplest version of a Poissonian model is one in which the energies and time-scales of the pulses and the background fraction are the same for all objects, regardless of their luminosities, redshifts, etc. We shall refer to this case as the Simple Poissonian model. This model makes two strong testable predictions: *(1)* the slope of the variability luminosity relation is $a = -1/2$, and *(2)* the energies and time-scales of the pulses are similar for objects from Seyfert galaxies to QSOs.

Our analysis of the SGP data showed that $a$ is dependent on the rather uncertain magnitude of the errors in $L$ and, to a lesser extent, on the variability index. Interestingly, the value $-0.5$ is nearly at the centre of the range of values resulting from different combinations of errors. We therefore regard these observations as being consistent with the Simple Poissonian model, though it is clear that we are not able to test definitively the $a = -0.5$ prediction with the data presently available.

On the assumption that the model is valid, the independent term $c$ can be used to estimate the typical energy of the pulses. Taking the logarithm of equation (9)

$$\log v_B = \frac{1}{2}\log f_{var} - \frac{1}{2}\log \frac{\overline{L_B}}{\overline{L_B^p}}$$

and comparing it to $f = ax + by + c$ computed for $y = 0$ (i.e. $z = 0$ and $\lambda = \lambda_B$), we derive

$$c = \frac{1}{2}\log\left(f_{var}\frac{\overline{L_B^p}}{10^{12}L_B^{\odot}}\right) \qquad (10)$$

The $v(\overline{L_B}, z)$ fits of the SGP sample yielded values of $c$ between $-1.1$ and $-0.8$ for those fits where $a \approx -0.5$, whilst the fits for the SF-based index $u(\overline{L_B}, z)$ resulted in a value of $c$ between $-1.1$ and $-0.7$, also for fits yielding $a \approx -0.5$. The theoretical prediction (equation 9) applies to the rms variability as measured from independent, uncorrelated points in the light curve, so in what follows we use the latter constraints for $c$. We then have that

$$6.3 \times 10^9 L_B^{\odot} \leq f_{var}\overline{L_B^p} \leq 4 \times 10^{10} L_B^{\odot} \qquad (11)$$

These limits, coupled with the constraint that $0.3 < f_{var} < 1$ yield limits on the pulse luminosity of

$$6.4 \times 10^9 L_B^{\odot} \leq \overline{L_B^p} \leq 1.3 \times 10^{11} L_B^{\odot} \qquad (12)$$

The value of $c$ does not by itself provide independent constraints on the average energy and time scale of the pulses. We can nevertheless use the time-scales derived from the SF analysis in Hook et al. (1994). Their study has shown that the *ensemble* SF of the QSOs in the SGP sample rises for $\Delta t \lesssim$ 2-3 yr and flattens off for larger lags, indicating that QSO variability is uncorrelated on longer time-scales (see also Cristiani et al. 1996). The pulse lifetime must therefore be in the neighbourhood of 2 yr, since the SF of a light curve resulting from the random superposition of many pulses is equal to the mean SF of the individual pulses. Assuming that the pulse lifetime is in the $1.5 < \overline{\tau_B^p} < 3$ yr range, then yields the following constraints on the pulse energy:

$$1.5 \times 10^{50}\ erg \leq \overline{E_B^p} \leq 6.3 \times 10^{51}\ erg \qquad (13)$$

This range incorporates the constraints in $c$, $f_{var}$ and $\overline{\tau_B^p}$, all of which are based on conservative estimates. The resulting limits on the pulse energy can therefore be regarded as a robust constraint. These limits are nevertheless cosmology dependent: $H_0 = 100$ km s$^{-1}$Mpc$^{-1}$ would result in energy limits 4 times smaller than those in (13).

### 7.3 Generalized Poissonian models

The simple model, while attractive and consistent with the data presently available, is not the only possible Poissonian model for QSO variability. More generally, one can envisage a situation in which the pulse properties (energy and time-scale) depend upon the global luminosity of the object. In this case the $v_B \propto \overline{L_B}^{-1/2}$ prediction would no longer be valid.

Consider a case where $\Lambda \equiv f_{var}\overline{L_B^p}$ depends on the luminosity of the object as $\Lambda(\overline{L_B}) \propto \overline{L_B}^{\gamma}$, the exponent $\gamma$ being determined by the physical process which links the properties of the individual pulses to those of the object as a whole. The variability luminosity relation then becomes $v_B \propto \overline{L_B}^{(\gamma-1)/2}$. The fit coefficient $c$ can now be used to calibrate $\Lambda(\overline{L_B})$, analogously to what was done for the Simple Poissonian model ($\gamma = 0$). Let us use as a reference point the luminosity $L_0$ of a 'typical' QSO, corresponding to $M_B = -26$, and write

$$\Lambda(\overline{L_B}) = \Lambda_0 \left(\frac{\overline{L_B}}{L_0}\right)^{\gamma} \qquad (14)$$

We shall restrict our discussion to three possible values of $\gamma$: 0.5, 0 and $-0.5$. These models would result in $v(L)$ slopes $a = -0.25$, $-0.5$ and $-0.75$, respectively. While the data analysis showed that we cannot as yet decide which of these possibilities is to be favoured, we can safely reject values of $a$ outside this range. For each value of $a$ we now estimate the corresponding value of the independent term $c$ from Fig. 8 and compute $\Lambda_0$. The results are listed in Table 3, where instead of $\Lambda_0$ we list $n_0 \equiv L_0/\Lambda_0$, i.e., the number of individual pulses which have to be added up to account for the luminosity of a $M_B = -26$ QSO. The actual mean number of pulses present at any moment is $N_0 = f_{var}^2 n_0$, which reduces to $n_0$ in the absence of a background component.

The values of $n_0$ obtained give us information on the product $f_{var}\overline{E_B^p}/\overline{\tau_B^p}$ at the calibration point $L_0$. In order to estimate typical pulse energies for the three models considered, let us take $f_{var} = 0.5$ and $\overline{\tau_B^p} = 2$ yr as fiducial values for an $L_0$ QSO. The resulting pulse energies are listed in Table 3, along with the required pulse rates, derived from $\nu = L_0 f_{var}/\overline{E_B^p}$. Note that the pulse energy for a Simple Poissonian model ($\gamma = 0$) is consistent with (13), as expected. The uncertainty in $c$ is $\pm 0.1$, which propagates to an uncertainty of 46 per cent for $n_0$, $\nu$ and $\overline{E_B^p}$.

| $\gamma$ | $a$ | $c$ | $n_0$ | $\nu/yr^{-1}$ | $\overline{E_B^p}/10^{51}\ erg$ |
|---|---|---|---|---|---|
| 0.5 | -0.25 | -0.75 | 61 | 7.6 | 3.8 |
| 0.0 | -0.50 | -0.90 | 231 | 28.9 | 1.0 |
| -0.5 | -0.75 | -1.10 | 1112 | 139.0 | 0.2 |

**Table 3.** Estimated parameters for generalized Poissonian models with $\gamma = 0.5$, 0 and $-0.5$. Pulse rates and energies correspond to a $M_B = -26$ QSO, and assume a background fraction of 50 per cent and a pulse lifetime of 2 yr.

### 7.4 Discussion

The most appealing aspect of the results presented above is their generality. The only assumption involved is that the pulses occur randomly in the light curve. While this leaves out cyclical/periodic processes, the analysis encompasses a large class of possible physical models.

Gravitational microlensing has recently been explored as an external source of variability in QSOs (Hawkins 1996). As the lensing events are probably independent, the whole process can be seen as Poissonian, though the luminosity of the QSO itself would act as a strong background. While this scenario predicts achromatic variability if all the emission arises from regions of the same scale-size, some chromaticism is expected if the continuum at differing wavelengths originates from different parts of an accretion disc (Wambsganss & Paczyński 1991, Alexander 1995). However, the size of the Broad Line Region in the canonical picture is too large to be magnified by the microlensing action of solar- and sub-solar mass stars at cosmological distances (Schneider & Wambsganss 1990), and no flux variations are expected in the emission lines. While this model is not ruled out as a general mechanism of variability in high-redshift QSOs (but see Alexander 1995), the variability properties of the majority of low-redshift QSOs clearly depart from the predictions of this scenario. Of the six well studied variable ($z < 0.3$) PG QSOs, four show line flux variations (Maoz et al. 1994). More generally, the higher variability of low-redshift, low-luminosity QSOs is contrary to the microlensing hypothesis, as the optical depth of microlensing objects would not be sufficient to produce the observed fluctuations (Wambsganss 1990, Lewis & Williams 1996).

One well-studied example of a Simple Poissonian model applied to AGN is the starburst model (Terlevich et al. 1992, Aretxaga & Terlevich 1994, Aretxaga et al. 1996). This model makes predictions for energies (3–6 $\times 10^{50}$ erg in the B band) and time-scales (1–4 yr) of the pulses, as well as for the background luminosity ($f_{bck} \approx 40$ per cent). The supernova rate goes from $\approx 3$ yr$^{-1}$ for an $M_B = -23$ QSO to 200 yr$^{-1}$ for a QSO with $M_B = -28$. These predictions are thoroughly consistent with the constraints derived in Section 7.2 (we refer the reader to Aretxaga & Terlevich 1994 and Aretxaga et al. 1996 for a discussion of this model). In the context of the super-massive black-hole scenario, AGN variability is traditionally associated with accretion instabilities (e.g. Rees 1984, Abramowicz 1991, Wallinder, Kato & Abramowicz 1992). If random accretion events are responsible for the variations observed in QSOs then they must produce B-band energies of the order of $10^{51}$ erg. For an efficiency of 10 per cent this yields a mass of 0.05 $M_\odot$ if 10 per cent of the photons produced come out in the B-band. Furthermore, about 30 such events must happen per year in an $M_B = -26$ QSO, so the accretion rate in the form of inhomogeneities must be about 1.5 $M_\odot\,yr^{-1}$.

Whilst Table 3 provides useful order of magnitude estimates of the rate and energy of the pulses in an $M_B = -26$ QSO, we cannot extrapolate them to other luminosities without detailed knowledge of individual dependences of $f_{var}$, $\overline{E_B^p}$ and $\overline{\tau_B^p}$ on the total luminosity. Such information would be very important for distinguishing between different scenarios. If, for instance, variability is associated with accretion onto a black-hole, one might perhaps associate the pulse duration with an orbital or viscous time-scale. If the accreted objects always have the same mass, one might expect the energy of the pulses to scale with the black-hole mass. Both time-scale and energy would then change as a function of luminosity.

Determining how the pulse properties scale with the object properties should therefore be a major goal of future variability studies. In principle, the pulse lifetime can be empirically determined from the SF of individual objects, but a precise measurement requires long and well sampled light curves. The evidence so far is that the time-scale as derived from the SF does not vary strongly as a function of luminosity (Hook et al. 1994, Cristiani et al. 1996). Pulse energies can only be directly measured if the events do not overlap. While this exercise can be tried in low luminosity AGN, in QSOs a large number of pulses are present at any time. This can be seen in Table 3: computing $N_0 \approx \nu\overline{\tau_B^p}$ yields at least 15 coexisting events in an $M_B = -26$ QSO. In such cases the pulse energy has to be derived indirectly, as above. Similarly, the background fraction cannot be determined when overlapping occurs, though it may be constrained through the method discussed in Section 7.1.

## 8 SUMMARY

The South Galactic Pole database was used to investigate the relationship between variability, luminosity and redshift in QSOs. This was done using parametric fits to the data in the 3D space defined by these three variables. Two variability indices were used: $v$, the net rms variability, and $u$, an index based on observations carried out more than 3 yr apart. This latter index avoids the problems related to time-dilation effects, but is subjected to larger errors than $v$. The variability-luminosity-redshift relation was fit with a $v \propto L^a(1+z)^b$ function, and similarly for $u$. This approach to the problem allows a careful examination of the effects of the errors involved. The main results of the data analysis can be summarized as follows:

- (i) We confirm Hook et al. 's (1994) result that variability is anticorrelated with luminosity in the SGP QSOs ($a < 0$ in all fits).
- (ii) We also found that variability increases with redshift ($b > 0$). This is in agreement with the results of Cristiani et al. (1996), but not with Hook et al.'s conclusion that the redshift correlation is a by-product of the coupling of $L$ and $z$. In particular, we found that the values of $b$ obtained for the SGP sample are consistent with those derived from the variability-wavelength anticorrelation observed in nearby AGN, indicating that this anticorrelation extends to luminous, high-$z$ objects (as first suggested by Giallongo et al. 1991).
- (iii) The logarithmic slope of the $v \times L$ relationship was found to lie in the $a = -0.7$ to $-0.2$ interval, depending on the magnitude of the errors involved. Similar slopes were obtained for the SF-based index $u(L)$. The uncertainties associated with sampling and particularly the k-correction do not allow a more precise determination of this slope.

The results of the data analysis were compared to theoretical predictions for Poissonian models, understood as systems in which the light curve is the result of a random superposition of pulses plus a possible background component. This comparison produced the following constraints which must be satisfied by physical models for the origin of the pulses:

- (i) The minima in the light curves were used to estimate the contribution of a constant 'background' source to the total B-band luminosity of the QSOs, yielding $f_{bck} < 70$ per cent.
- (ii) We explored the hypothesis that QSO variability can be explained in a Simple Poissonian model, in which the energy and time-scales of the pulses are the same for all QSOs independent of their luminosity. This model predicts a 'one over square root of $N$' relation between the net rms variability and the luminosity. We were able to put hard limits on the typical B-band luminosity of the pulses. These can then be coupled with the $1.5 < \tau_B^p < 3$ yr limits on the time-scale of the pulses, as derived from the analysis of the ensemble SF, to constrain the B-band energy of the pulses between $1.5 \times 10^{50}$ and $6.3 \times 10^{51}$ erg.
- (iii) Generalized Poissonian models, in which the pulse properties scale with the QSO luminosity, were also investigated. Models in which the pulse and object luminosities are related by $\overline{L_B^p} \propto \overline{L_B}^\gamma$ are also consistent with the SGP data provided $-0.5 \le \gamma \le 0.5$. For an $M_B = -26$ QSO the pulse rates and B-band energies required in such models were all found to be within an order of magnitude of 30 $yr^{-1}$ and $10^{51}$ erg respectively.

Future observations should concentrate on reducing the errors both in variability and in luminosity, and on improving the coverage of the $L$-$z$ plane. Spectrophotometry and/or multi-colour monitoring of QSOs could in principle provide more stringent tests of the wavelength dependence of variability. This would automatically remove the wavelength effect, something which our simple $v(\lambda) \propto \lambda^{-b}$ law could only do in an approximate way. Residual correlations with redshift, if any, would then have to be explained in another way (for instance, as due to cosmic age). It is also important to encourage the continuation of monitoring programs such as that of Hook et al. (1994), since their wide time baseline provides the type of long term information which we cannot hope to obtain in the near future. Also, as the length of the light curves increases the errors associated with sampling effects should decrease. It would also be interesting to study the dependence of variability on properties other than $L$ and $z$. For instance, radio-loud objects seem to vary more than their radio-quiet counterparts (Pica & Smith 1983, Maoz et al. 1994). Radio observations of the SGP objects would be important to verify this property and to better study the $v$-$L$-$z$ relationship in these two families of AGN.

Until now variability studies of high- and low-luminosity AGN have been largely treated as separate topics. In order to further our understanding of AGN, it will be of great importance to link the variability properties of QSOs to those of Seyferts and LINERs. Only then will we be able to address fundamental questions such as: Do low luminosity AGN follow the same variability-luminosity-wavelength relation as QSOs? Is the process Poissonian? What is the rate, energy and time-scales of the pulses and how do they vary from object to object? Answering these questions would narrow considerably the range of physical models for AGN variability.

**Acknowledgments:** We are greatly indebted to Isobel Hook, Richard McMahon, Mike Irwin and Brian Boyle for making the SGP data available to us. Discussions with Mike Irwin, Geraint Lewis, Richard McMahon, Onno Pols and Laerte Sodré are also acknowledged. The hospitality of ESO (Santiago), where this work was completed, was much appreciated. The work of RCF was supported by CAPES/Brazil under grant 417/90-5. IA's work is supported by the EEC HCM fellowship ERBCHBICT941023.